\documentstyle[prl,aps,eqsecnum,epsfig,multicol]{revtex}

\newcommand {\ppbar}    {\mbox{$p\overline{p}$}}

\newcommand {\Rinv}     {\mbox{$R_{\text{inv}}$}}

\newcommand {\Rs}       {\mbox{$R_{\text{side}}$}}
\newcommand {\Ro}       {\mbox{$R_{\text{out}}$}}
\newcommand {\Rl}       {\mbox{$R_{\text{long}}$}}

\newcommand {\Rg}       {\mbox{$R_{\text{geom}}$}}

\newcommand {\qs}       {\mbox{$q_{\text{side}}$}}
\newcommand {\qo}       {\mbox{$q_{\text{out}}$}}
\newcommand {\ql}       {\mbox{$q_{\text{long}}$}}

\newcommand {\kT}       {\mbox{$k_T$}}
\newcommand {\mT}       {\mbox{$m_T$}}

\newcommand {\bT}       {\mbox{$\beta_{\text T}$}}

\newcommand {\snn}      {\mbox{$\sqrt{s_{\text{NN}}}$}}

\newcommand {\RoPCMS}   {\mbox{$R_{\text{out}}^{\text{PCMS}}$}}

\def\PLB{{Phys. Lett.}~{\bf B}}
\def\PRP{Phys. Rep.\ }
\def\PRL{Phys. Rev. Lett.\ }
\def\PRD{{Phys. Rev.}~{\bf D}}
\def\PRC{{Phys. Rev.}~{\bf C}}

\def\EPJC{{Eur. Phys. J.}~{\bf C}}

\begin{document}
\draft

\title{Transverse mass dependence of two-pion correlations \\
in Au+Au collisions at $\snn = 130$~GeV} 

\author{
K.~Adcox,$^{40}$
S.{\,}S.~Adler,$^{3}$
N.{\,}N.~Ajitanand,$^{27}$
Y.~Akiba,$^{14}$
J.~Alexander,$^{27}$
L.~Aphecetche,$^{34}$
Y.~Arai,$^{14}$
S.{\,}
% ...
\\(PHENIX Collaboration)\\
}
\address{
% ...
$^{3}$Brookhaven National Laboratory, Upton, NY 11973-5000, USA\\
% ...
$^{14}$KEK, High Energy Accelerator Research Organization, Tsukuba-shi, Ibaraki-ken 305-0801, Japan\\
% ...
$^{27}$Chemistry Department, State University of New York - Stony Brook, Stony Brook, NY 11794, USA\\
% ...
$^{34}$SUBATECH (Ecole des Mines de Nantes, IN2P3/CNRS, Universite de Nantes) BP 20722 - 44307, Nantes-cedex 3, France\\
% ...
$^{40}$Vanderbilt University, Nashville, TN 37235, USA\\
% ...
}

%

\date{\today}        
\maketitle
\begin{abstract}
Two-pion correlations in $\snn=130$~GeV Au+Au collisions at RHIC
have been measured over a broad range of pair transverse momentum
$k_T$ by the PHENIX experiment at RHIC.  The \kT~dependent transverse
radii are similar to results from heavy ion collisions at $\snn$ =
4.1, 4.9, and 17.3 GeV, whereas the longitudinal radius increases
monotonically with beam energy.  The ratio of the outwards to
sidewards transverse radii (\Ro/\Rs) is consistent with unity and
independent of \kT.
\end{abstract}

\pacs{PACS numbers: 25.75.Dw}

\begin{multicols}{2}   
\narrowtext            

The influence of Bose-Einstein statistics on the correlation of
identical charged pions at low relative momentum was first used to
probe the space-time structure of pion emission in
\ppbar~annihilations~\cite{gol60} and has subsequently been applied to
relativistic heavy-ion collisions from the Bevalac to
RHIC~\cite{zaj84,lis00,soltz01,na4400,wa98,starprl01}
(see~\cite{wie99} for a recent review), and to a wide range of systems
including $e^+e^-$ annihilations~\cite{epem}.  
The correlation function is defined as the ratio of the two-particle
probability distribution to the product of the single particle
distributions.  For a static source with no final state interactions,
it is related to the Fourier transform with respect to
${\bf q} = {\bf p_1} - {\bf p_2}$ of the source distribution $\rho
({\bf r})$,
${P({\bf p_1},{\bf p_2})}/{P({\bf p_1})P({\bf p_2})} = 
1 + |\tilde{\rho}({\bf q})|^2$~\cite{gol60}.
If the source is parameterized as a multi-dimensional Gaussian, the
enhancement in the correlation function is a Gaussian, and the
Gaussian widths are each inversely proportional to the source
dimensions in the canonically conjugate spatial variables.  The
extracted source dimensions are commonly referred to as HBT radii,
after a similar technique developed by Hanbury-Brown and Twiss to
measure stellar radii~\cite{hbt54}.
For dynamic sources, such as rapidly expanding sources in
heavy-ion collisions, the correlation function measures ``lengths of
homogeneity'', or the relative separations of the pions with low
relative momentum.  This leads to source radii which depend strongly
on \kT, the mean transverse momentum of the pion
pair~\cite{pra84,pra86,mak88,cha95ext,fie95,wie96}.  If the
dynamics are correctly modeled, 
then both the source geometry and rate of expansion
can be deduced by measuring the \kT~dependence of the
radii.  The existence of a connection between HBT radii and heavy-ion
source geometry is established by the dependence of the radii
on system size~\cite{bar86}, centrality~\cite{soltz01,na4400}, and
reaction plane~\cite{lis00}.  Interest in Bose-Einstein correlations
in heavy-ion collisions is driven by the expectation that HBT radii
are sensitive to the large and/or long-lived sources which may
accompany a QCD phase transition~\cite{pra86,ber89}.  Recent
calculations predict that the greatest sensitivity to a long-lived
source will come from measurements of the correlation function at high
\kT~($\geq 0.3$~GeV/c)~\cite{ris96,soff}.  

We present new measurements from the PHENIX experiment on two-pion
correlations in Au+Au collisions at $\snn=130$~GeV in the region
$|\eta|<0.35$, $0.2<\kT<1.0$ GeV/c, significantly extending previous
measurements by STAR~\cite{starprl01} up to a mean-\kT~0.63~GeV/c.  The data
are compared to theoretical predictions for RHIC and to HBT radii from
lower energy collisions at the SPS and AGS.  The \kT~dependence of the
transverse radii is used to extract a geometric transverse radius.

The PHENIX experiment has been described in detail elsewhere
\cite{phenix_description,hamagaki}.  For this analysis we utilize a
subset of the detectors in PHENIX.  We use
the hadronic particle identification capabilities present in the west arm
of the PHENIX spectrometer perpendicular to the beam direction
\cite{hamagaki} with polar and azimuthal ranges of $|\eta|< .35$ and
$\pi/4$, respectively, during its first year of running.  In this
analysis, the vertex is determined with a zero degree calorimeter
(ZDC) and a pair of Cerenkov beam-beam counters (BBC).  Pattern recognition
and momentum reconstruction rely on a drift chamber and a pad chamber
which occupy the region between 2.0 and 2.5 meters from the beam axis.
The momentum resolution from these detectors is $\delta p/p = 0.6\%
\oplus 3.6\%p$.  Particle velocity is determined from the differential
time measurements of the BBC and the electromagnetic calorimeter
(EMC)~\cite{swqm01}, with a combined rms resolution of 600 ps, coupled
with the path length determined from pattern recognition.  The
momentum determination and particle identification method are similar
to~\cite{hadron_paper}, except that the time of flight is measured by
the EMC.  A pion is defined as being within 1.5 standard deviations of
the pion mass-squared peak but at least 2.5 standard deviations away
from the kaon peak.  After applying inter-detector association cuts the
background from mis-associated EMC hits is $\sim 10\%$ as determined
by a hit randomization technique.  This background does not
significantly distort the extracted radius in the correlation
measurements, although it reduces the measured correlation strength
($\lambda$).  We did not correct for this background in our
correlation analysis.

A total of 493K events in the most central 30\% of the cross section
survive all offline cuts. This sample contains 3.1 million $\pi^+$
pairs and 3.3 million $\pi^-$ pairs in the analysis, and has a mean
centrality of 10\%.

The pion correlation function is determined from pairs of identical
pions.  The normalized probability of detecting two particles with
relative momentum ${\bf q}={\bf p_1}-{\bf p_2}$ and average momentum
${\bf k}=({\bf p_1}+{\bf p_2})/2$ is determined experimentally by the
ratio of pairs from the same event ($A$) with those from different
events ($B$): $C_2({\bf q},{\bf k}) = A({\bf q},{\bf k})/B({\bf
q},{\bf k})$.  Pairs of particles within 2 cm of each other in the
drift chamber are eliminated from the analysis in both the real and
background samples.  Pairs that share the same EMC cluster are also
removed from both samples.  Finally, all pairs in the mixed background
sample are required to be from events with a reconstructed BBC
collision vertex within 1cm of each other.

We correct for the Coulomb interaction of the pairs in the correlation
function by parameterizing the source as a Gaussian distribution in
the pair center-of-mass frame and performing an iterative procedure
\cite{baker} which accounts for the finite resolution of the detector.
This procedure applied to the distribution of $\pi^+$-$\pi^-$ pairs is
in agreement with the data, although the statistics in the Run-1
opposite-signed analysis are not sufficient to independently determine
the required Coulomb correction.  Systematic studies of the Coulomb
correction which vary both radius and magnitude within reasonable
constraints produce variations in the final radii which never exceed
0.25 fm.

The relative momenta are projected into the variables $\ql$, along the
beam direction, $\qo$, parallel to the transverse momentum of the pair
${\bf \kT} = \frac{1}{2} ({\bf p_{T1}}+{\bf p_{T2}})$, and $\qs$,
perpendicular to $\ql$ and $\qo$~\cite{pra84,ber89}.  These variables
are calculated in the longitudinal co-moving system (LCMS), obtained
by a longitudinal boost from the lab frame to the frame in which the
longitudinal pair velocity vanishes.  This frame is commonly used for
sources expected to be invariant under longitudinal boosts~\cite{bj}.

\begin{figure}
\centerline{\epsfig{file=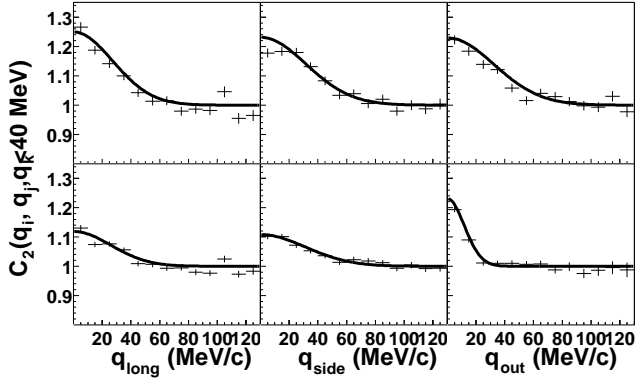,width=1.0\linewidth}}
\caption[]{The three dimensional correlation function for $\pi^-$
pairs versus \ql,\qs, and \qo~in both the LCMS frame (top) and pair
center-of-mass frame (bottom).  The data are plotted versus one
momentum difference variable while requiring the other two to be less
than 40 MeV/c.  The lines correspond to the fit to the entire
distribution.}
\label{f:fig1} 
\end{figure}

The fully corrected correlation function for $\pi^-$ pairs is shown
in the top panels of Fig.~\ref{f:fig1}; the large $q$ region of the
correlation function has been normalized to 1 in the plots.  The data
are fit to a Gaussian parameterization of the source using a MINUIT
based log-likelihood method~\cite{soltz01}.
\begin{equation}
C_2 = 1 + \lambda \exp(-\Rl^2\ql^2 - \Rs^2\qs^2 - \Ro^2\qo^2)
\end{equation}
where $\Rl$, $\Rs$, and $\Ro$ are the conjugate variables to $\ql$,
$\qs$, and $\qo$, respectively.
Errors quoted in the tables and figures are statistical
only.  Systematic errors come mainly from the Coulomb correction and
dependence of the results on the two-track distance cuts.  The
combined systematic error for these effects, estimated by varying the
cuts and corrections within reasonable bounds, is 8\% for $\Rl$, and
$\Rs$, and 4\% for $\Ro$. 
The systematic error from residual correlations in the event-mixed
background \cite{zaj84} is 2\%, 
yielding a total systematic error of $\sim$8\% for $\Rl$ and $\Rs$ and
$\sim$4.5\% for $\Ro$.  

\begin{figure}
\centerline{\epsfig{file=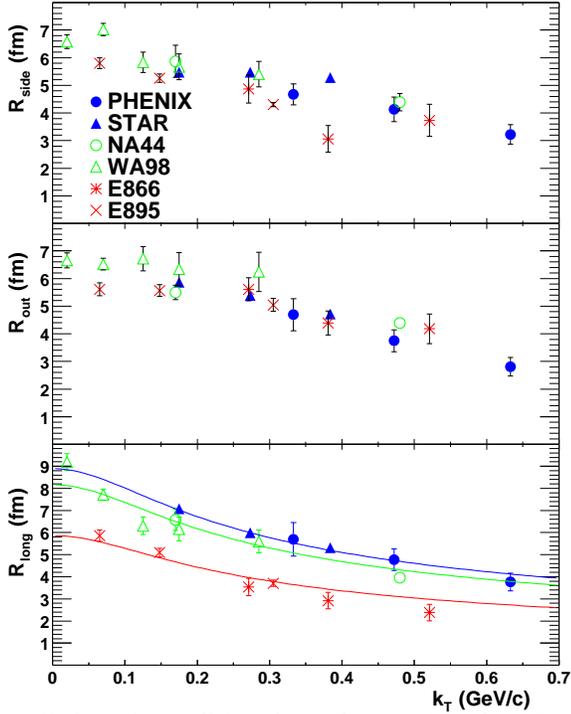,width=1.0\linewidth}}
\caption[]{HBT radii for pion pairs as a function of $k_T$ 
measured at mid-rapidity for various energies from E895 ($\snn$ =
4.1~GeV), E866 ($\snn$=4.9 GeV), NA44, WA98 ($\snn$=17.3 GeV), STAR, and
PHENIX ($\snn$=130 GeV).  The bottom plot includes fits to
$A/\sqrt{\mT}$ for each energy region.  The data are for $\pi^-$
results except for the NA44 results, which are for $\pi^+$.}
\label{f:roots}
\end{figure}

The data set is subdivided into three $\kT$ bins of equivalent
statistics in order to study the momentum dependence of the
correlation function.  In Fig.~\ref{f:roots}, the radii for
$\pi^-$~pairs are shown to agree within statistical and systematic
errors with previous measurements for overlapping $\kT$ bins at this
energy for the 12\% most central events.  For STAR,
the mean pair centrality can be approximated by the
geometric mean of 8\%, which is slightly more central than the mean pair
centrality of 10\% for the PHENIX data.  This figure also shows $\kT$
dependent radii for mid-rapidity pions from central collisions for
$\snn=17.3$~GeV Pb+Pb~\cite{wa98,na44} and for
$\snn=4.9$~and~4.1~GeV~Au+Au~\cite{lis00,soltz01}.  For the transverse
radii, \Ro~and~\Rs, the variation with collision energy is generally
smaller than the statistical and systematic errors of the individual
data points.  There is no evidence for a change in the
low-\kT~extrapolation of \Rs~with increasing \snn~which would indicate
a larger geometric source at higher energy.  Nor is any change evident
in \Ro~relative to \Rs~at high-\kT, indicating a longer-lived source.
This result is surprising given the factor of $\sim$3 change in the
total charged particle multiplicity per unit rapidity at
mid-rapidity\cite{mult}.  Only \Rl~exhibits a significant variation
with collision energy.  To quantify this difference, we fit the $\Rl$
dependence to $A/\sqrt{\mT}$~\cite{mak88,wie96,rlnote} for the three
sets of beam energies.  The results are overlayed with the data in the
bottom panel of Fig.~\ref{f:roots} and yield $A = 3.32 \pm 0.03$,
$3.05 \pm 0.06$, and $2.19~\pm~0.05$~fm$\cdot$GeV$^\frac{1}{2}$ for
$\snn=130$, 17.3, and 4.9/4.1~GeV respectively.

\begin{figure}
\centerline{\epsfig{file=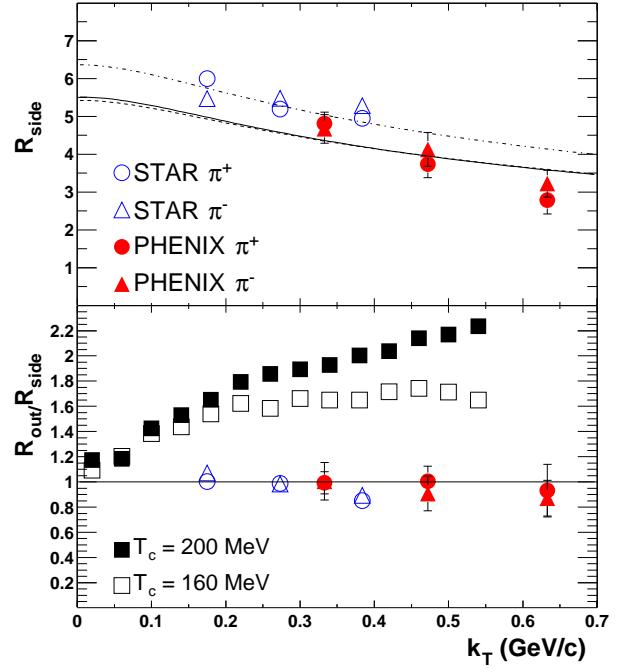,width=1.0\linewidth}}
\caption[]{The top panel shows the measured $\Rs$ from identical pions
for STAR and PHENIX.  The line solid line is a fit of
Eq.~\ref{eq:rswied} to the PHENIX data, and the dashed line is the
same fit for Eq.~\ref{eq:rschap}.  The dot-dashed line is a fit of
Eq.~\ref{eq:rswied} to the STAR data.  The bottom panel shows the
ratio $\Ro/\Rs$ as a function of $\kT$ overlayed with theoretical
predictions for a phase transition for two critical temperatures.}
\label{f:rsandro}
\end{figure}

Although a finite emission duration contributes to \Ro~but not to \Rs,
dynamical correlations affect the two radii differently.  A
quantitative determination of the source lifetime can only be
performed in the context of a dynamical model.  The lower panel of
fig.~\ref{f:rsandro} shows the \kT~dependence of the ratio
\Ro/\Rs~for PHENIX and STAR along with recent calculations for a 
thermalized source which undergoes a first order phase transition at
critical temperatures ($T_c$) of 160~and~200~MeV~\cite{soff}.  The
rise in \Ro/\Rs~which comes predominantly from a hadronic
re-scattering phase is not present in the data, and the values
of 1.6 ($T_c=160$~MeV) and 2.2~($T_c=200$~MeV) at high \kT~are excluded.

An additional consequence of strong dynamics occurs for sources in
which the transverse expansion is relativistic.  In this case,
\Ro~measured in the LCMS frame is Lorentz contracted by the $\gamma$
of the pion source velocity along the direction of
\qo~\cite{pra90,zaj93}.  Current Lorentz invariant formulations of the
correlation function~\cite{yk78,cha95ext} are insufficient to
determine the source velocity due to transverse expansion, however,
the pair center-of-mass system (PCMS) can be used to provide an upper
limit on \Ro~\cite{pcmsnote}.  The correlation function
for $\pi^-$ pairs in the PCMS frame is shown in the bottom panels of 
Fig.~\ref{f:fig1}, and fit results for 
\RoPCMS~are listed in Table~\ref{t:table1}.  As expected, 
\Rs~and~\Rl~are equal to the corresponding LCMS parameters within errors.

Two analytic expressions have been used to describe \Rs~as a function
of $\mT=\sqrt{\kT^2+m^2_{\pi}}$ for a transversely expanding source,
\begin{eqnarray}
{\Rs^2(\mT)} & = & \frac{\Rg^2}{1+\beta^2_f \left( \frac{\mT}{T} \right)},
\label{eq:rschap} \\
{\Rs^2(\mT)} & = &
\frac{\Rg^2}{1+\eta^2_f\left(\frac{1}{2}+\frac{\mT}{T}\right)}. 
\label{eq:rswied}
\end{eqnarray}
Eq.~\ref{eq:rschap} is a first order approximation in $\frac{T}{\mT}$
for a longitudinally boost invariant source with finite temperature,
$T$, and expansion velocity, $\bT=\beta_f \rho/\Rg$, where \Rg~is the
Gaussian transverse radius~\cite{cha95ext}.
Eq.~\ref{eq:rswied} includes an additional term in the approximation
and the linear transverse expansion velocity is replaced by a
transverse rapidity, $\eta_T=\eta_f \rho/\Rg$~\cite{wie96}.  For a
transverse surface rapidity of $\eta_f=0.85$ ($\beta_f=0.69$) and
$T=125$~MeV~\cite{ppg009,flownote}, a fit of Eq.~\ref{eq:rswied} to
the PHENIX \Rs~\mT~dependence yields, $\Rg=8.1 \pm 0.3$~fm with a
$\chi^2/{\text{dof}}=9.6/6$.  To assess systematic errors the PHENIX
data are also fit to Eq.~\ref{eq:rschap}, yielding $\Rg=6.7 \pm 0.2$~fm
and $\chi^2/{\text{dof}}=9.1/6$, and the STAR data are fit to
Eq.~\ref{eq:rswied}, yielding $\Rg=9.4 \pm 0.1$~fm with
$\chi^2/{\text{dof}}=21/6$.  These fits are shown in the top panel of
Fig.~\ref{f:rsandro}.  All values of \Rg~are significantly larger than
the comparable 1D rms radius for a Au nucleus~\cite{hahn} of
$\sqrt{\frac{1}{3}}\cdot\sqrt{\frac{3}{5}}\cdot 6.87 = 3.07$~fm.

In conclusion, we have extended the measurement of two particle
correlations for Au+Au collisions at $\snn$= 130 GeV to $<\kT>$ =
0.63~GeV/c using the PHENIX detector at RHIC.  Values of \RoPCMS~are
used to constrain the Lorentz effects for a relativistic transverse
expansion.  Fitting \Rs(\kT) to two analytic expressions for an
expanding source yields a transverse geometric radius that is much
larger than the comparable radius for Au.  We find that
\Rl(\kT)~increases monotonically with collision energy, yet no energy
dependence is discernible in the \kT~dependence of \Ro~and \Rs, and
the ratio, \Ro/\Rs, is consistent with unity and independent of \kT.
The results for the transverse radii are contrary to common
expectations for a first order phase transition in Au+Au collisions at
these energies, as demonstrated by the comparison to a typical
hydrodynamic model with hadronic-rescattering.  Therefore, we conclude
that current concepts regarding the space-time evolution of the pion
source inferred from two-pion correlations in Au+Au collisions at RHIC
will need to be revised.

%
%
%
%
%
%
%
%


We thank the staff of the Collider-Accelerator and Physics Departments at
BNL for their vital contributions.  We acknowledge support from the
Department of Energy and NSF (U.S.A.), MEXT and JSPS (Japan), RAS,
RMAE, and RMS (Russia), BMBF, DAAD, and AvH (Germany), VR and KAW
(Sweden), MIST and NSERC (Canada), CNPq and FAPESP (Brazil), IN2P3/CNRS
(France), DAE and DST (India), KRF and CHEP (Korea), the U.S. CRDF for 
the FSU, and the US-Israel BSF.





\newpage   

\begin{table} 
\caption{The \kT~dependencies of the $\pi^+$ and $\pi^-$ radii in
the LCMS and PCMS frames. All momenta are in MeV and
all radii are in fm.  The errors are statistical only. \label{t:table1}}
\begin{tabular}[]{ccccc}

 & $\kT$ (MeV)           & $200-400$ & $400-550$ & $550-1000$ \\

 & $\langle \kT \rangle$ & $333$     & $472$     & $633$      \\ \hline

 & 
$\Rinv$ &
$6.74 \pm 0.31$ &
$6.42 \pm 0.46$ &
$3.46 \pm 0.46$ \\

 & 
$\lambda_{LCMS}$   &
$0.423 \pm 0.037$ & 
$0.389 \pm 0.039$ & 
$0.287 \pm 0.048$ \\ 

$\pi^+$ &
$\Rl$ &
$6.01 \pm 0.45$ & 
$4.76 \pm 0.35$ & 
$2.97 \pm 0.38$ \\

 & 
$\Rs$ &
$4.81 \pm 0.30$ & 
$3.74 \pm 0.36$ & 
$2.79 \pm 0.37$ \\

 & 
$\Ro$ &
$4.78 \pm 0.30$ & 
$3.76 \pm 0.26$ & 
$2.59 \pm 0.46$ \\

 & 
$\RoPCMS$ &
$11.35 \pm 0.69$ &
$12.20 \pm 1.02$ &
$8.60 \pm 1.13$ \\ \hline

 & 
$\Rinv$ &
$6.00 \pm 0.30$ &
$5.96 \pm 0.41$ &
$4.58 \pm 0.48$ \\

 & 
$\lambda_{LCMS}$   &
$0.431 \pm 0.079$ & 
$0.405 \pm 0.067$ & 
$0.353 \pm 0.062$ \\ 

$\pi^-$ &
$\Rl$ &
$5.69 \pm 0.76$ & 
$4.77 \pm 0.49$ & 
$3.76 \pm 0.41$ \\

 & 
$\Rs$ &
$4.67 \pm 0.38$ & 
$4.13 \pm 0.45$ & 
$3.22 \pm 0.35$ \\

 & 
$\Ro$ &
$4.69 \pm 0.58$ & 
$3.75 \pm 0.40$ & 
$2.81 \pm 0.34$ \\ 

 & 
$\RoPCMS$ &
$11.27 \pm 0.72$ &
$12.42 \pm 1.18$ &
$11.89 \pm 1.73$ \\ 
\end{tabular} 
\end{table}

\end{multicols}    


\begin{references}
\bibitem[*]{Deceased}Deceased     
\bibitem[\dagger]{non-par}  
	Not a participating Institution.

\bibitem{gol60}{G. Goldhaber {\em et al.}, Phys. Rev. {\bf 120}, 300
(1960).}

\bibitem{zaj84}{W.A. Zajc {\em et al.}, \PRC {\bf 29}, 2173 (1984).}

\bibitem{lis00}{M. Lisa {\em et al.}, \PRL {\bf 84}, 2798 (2000).}

\bibitem{soltz01}{R. Soltz, M.D.~Baker, J.H.~Lee, for E802 Coll.,
Nucl.~Phys.~{\bf A661}, 439 (1999), L.~Ahle {\it et al.}, to be
submitted to \PRC, (2002).}

\bibitem{na4400}{I.G. Bearden {\it et al.}, Eur. Jour. Phys. {\bf
C18}, 317 (2000).}

\bibitem{wa98}{M.M.~Aggarwal {\it et al.}, Eur. Phys. J. {\bf C16}, 445 (2000).}

\bibitem{starprl01}{C. Adler {\em et al.}, \PRL {\bf 87}, 082301 (2001).}

\bibitem{wie99}{U. Wiedemann and U. Heinz, \PRP {\bf 319}, 145, (1999).}

\bibitem{epem}{G.~Abbiendi {\it et al.}, \EPJC {\bf 16},
423, (2000).}

\bibitem{hbt54}{R.~Hanbury-Brown and R.~Twiss, Phil.~Mag. {\bf 45},
663, (1954).}

\bibitem{pra84}{S. Pratt, \PRL {\bf 53}, 1219 (1984).}

\bibitem{pra86}{S. Pratt, \PRD {\bf 33}, 1314 (1986).}

\bibitem{mak88}{A.N. Makhlin and Y.M. Sinyukov, Z. Phys. {\bf C39},
69 (1988).}

\bibitem{cha95ext}{S. Chapman, J.R. Nix, U. Heinz, \PRC {\bf 52}, 2694
(1995).}

\bibitem{fie95}{D.E. Fields {\it et al.}, \PRC {\bf 52}, 986 (1995).}

\bibitem{wie96}{U. Wiedemann, P. Scotto, U. Heinz, \PRC {\bf 53}, 918
(1996).}

\bibitem{bar86} {J.~Bartke, \PLB {\bf 174}, 32 (1986).}

\bibitem{ber89}{G. Bertsch and G.E. Brown, \PRC {\bf 40}, 1830 (1989).}

\bibitem{ris96}{D. Rischke and M. Gyulassy, Nucl. Phys. {\bf A608},
479 (1996).}

\bibitem{soff}{S. Soff, S.A. Bass, and A. Dumitru, \PRL
{\bf 86}, 3981 (2001).}

\bibitem{phenix_description}{D.P.~Morrison, Nucl. Phys. {\bf A638},
565c (1998), N. Saito, {\it ibid}, 575c (1998).}

\bibitem{hamagaki}{H.~Hamagaki for PHENIX Coll., Nucl.~Phys.~{\bf A698}, (2002).}

\bibitem{swqm01}{S.~White for PHENIX Coll., Nucl.~Phys.~{\bf A698}, (2002).}

\bibitem{hadron_paper}{K.~Adcox {\it et al.} nucl-ex/0112006,
submitted to \PRL (2001).}

\bibitem{baker} {M.D.~Baker, Nucl. Phys. {\bf A} {\bf 610}, 213c (1996).}

\bibitem{bj}{J.D. Bjorken, \PRD {\bf 27}, 140 (1983); F.~Cooper,
G.~Frye, E.~Schonberg, \PRD {\bf 11}, 192 (1974).}

\bibitem{na44}{I.G.~ Bearden {\it et al.}, NA44 Collaboration, \PRC
{\bf 58}, 1656 (1998).}

\bibitem{mult}{L. Ahle {\it et al.}, \PRC {\bf 57}, R466 (1998);
D.~B.~Back {\it et al.}, \PRL {\bf 85}, 3100 (2000).}

\bibitem{rlnote}{This functional form is motivated by an approximation
in $T/\mT$ in which $A=\tau_0 T$, where $\tau_0$ is the proper time of
hadronization.} 

\bibitem{pra90}{S.~Pratt, T.~Csorgo, and J.~Zimanyi, \PRC
{\bf 42}, 2646 (1990).}

\bibitem{zaj93}{``Particle Production in Highly Excited Matter'', Gutbrod
and Rafelski ed., Plenum Press, 1993, p435.}

\bibitem{yk78}{F.~Yano and S.~Koonin, \PLB {\bf 78}, 556 (1978).}

\bibitem{pcmsnote}{Here the radii are Lorentz {\em
extended} by $\gamma_s$ measured in the PCMS frame.  For a
toy model of an azimuthally symmetric source in motion but not
expanding, $\RoPCMS=\gamma_s \sqrt{\Rs^2 + \beta_s^2 \tau^2}$.}

\bibitem{ppg009}{K.~Adcox {\it et al.}, to be submitted to \PRC, (2002).}

\bibitem{flownote}{These values are taken from fits to the single
particle spectra for the 5-15\% centrality bin for a linear velocity
profile in a hard sphere.  In the same reference it is shown that for
a linear transverse rapidity in a Gaussian source these values vary by
less than 2\%.}

\bibitem{hahn}{B. Hahn, D.G.~Ravenhall, and R. Hofstadter,
Phys. Rev {\bf 101}, 1131 (1956).}

\end{references}
\end{document}